\documentclass [12pt]{article}
\usepackage{amssymb,amsfonts,latexsym,amsthm,times}
\usepackage{fancyhdr}
\usepackage{color}
\usepackage[english]{babel}
\usepackage[pdftex]{graphicx}
\usepackage{amsmath}
\usepackage{amscd}
\usepackage{verbatim}

\theoremstyle{definition}

\theoremstyle{remark}

\newcommand{\be}{\begin{equation}}
\newcommand{\ee}{\end{equation}}
\newcommand{\bdm}{\begin{displaymath}}
\newcommand{\edm}{\end{displaymath}}
\newcommand{\bea}{\begin{eqnarray*}}
\newcommand{\eea}{\end{eqnarray*}}
\newcommand{\bean}{\begin{eqnarray}}
\newcommand{\eean}{\end{eqnarray}}

\begin{document}


\title{A Model of the Ebola Epidemics in West Africa \\ Incorporating Age of Infection}

\author{    G.F. Webb$^{}$$^{}$\thanks{Corresponding author. Email: glenn.f.webb@vanderbilt.edu
\vspace{6pt}} and C.J. Browne$^{}$\\\vspace{6pt}  $^{}${\em{Vanderbilt University, Nashville, TN USA}}}

\maketitle

\begin{abstract}
A model of an Ebola epidemic is developed with infected individuals structured according to disease age. The transmission of the infection is tracked by disease age through an initial incubation (exposed) phase, followed by an infectious phase with variable transmission infectiousness. The removal of infected individuals is dependent on disease age, with three types of removal rates: (1) removal due to hospitalization (isolation), (2) removal due to mortality separate from hospitalization, and (3) removal due to recovery separate from hospitalization. The model is applied to the  Ebola epidemics in Sierra Leone and Guinea. Model simulations indicate that successive stages of increased and earlier  hospitalization of cases have resulted in mitigation of the epidemics. 
\end{abstract}
\vspace{,2in}

\section{Introduction}

The 2014-2015 Ebola epidemics in West Africa have been mathematically modeled in many studies with attention to various aspects of the disease epidemiology (References). In this study we focus attention on the variability of infectiousness of infected individuals throughout their disease course. It is recognized that the period of infectiousness follows a period of incubation lasting from 2 to 21 days, begins with the presentation of symptoms, and that the level of infection transmission  coincides with the severity of symptoms \cite{WHO2}. Our analysis of infection transmission is based on the disease age of an infected individual. Here disease age means the time since infection acquisition, which is assigned disease age $0$.  We assume that the disease age distribution of the level of infectiousness is a Gaussian distribution, rising gradually, peaking, and falling gradually for survivors. The value of tracking disease course by disease age is that removal of infected individuals from the epidemic population can be tied to their level of infectiousness, and earlier removal can have significant benefit for mitigation of the epidemic. 

The Ebola epidemics in West Africa have developed with rapid increase in the cumulative number of cases from the spring of 2014 to the present. The most recent WHO data indicate that the epidemics may be subsiding \cite{WHO1} . In this study we analyze the dynamics of the epidemics in Sierra Leone and Guinea with a disease age model. Our goal is to connect the removal rates of infected individuals to the epidemic outcomes. Our simulations indicate that the mitigation of the epidemics is tied to an increased level and earlier hospitalization (isolation) of infectious individuals as the epidemics unfolded. This enhanced hospitalization of identified cases can be tied to increased contact tracing, social awareness, and public health policies. The measure of these epidemics is tracked by the net reproductive values $R_0$ derived from the models. Our simulations show that for both Sierra Leone and Guinea  $R_0$ has decreased significantly to levels which indicate that the epidemics will subside in 2015. We fit our disease age structured model simulations to cumulative case data and extend our simulations forward in time  to project the epidemic levels into 2015. 

\section{The Ebola Model with Age of Infection}

The model consists of the epidemic population divided into susceptible $S(t)$,  infected $I(t)$, and removed classes $R(t)$ at time $t$, with the key feature that infected individuals are structured by a continuous variable $a$ corresponding to time since acquisition. The age of infection is divided into an early stage corresponding to an incubation period, and intermediate stage corresponding to an infectious period, and a late stage corresponding to a recovering period, as determined by the disease age of infected individuals. 

\begin{enumerate}
 \item Initially there are $S_0$ susceptible individuals with a small number of infected individuals. The results are valid for any number of initial susceptibles with a given parameterization. 
  \item The disease phases of  an infected individual are governed by a Gaussian distribution of disease age $a$.  The Gaussian distribution has mean $\alpha_m$, standard deviation $\alpha_{sd}$, and is multiplied by a scaling factor 
$\alpha_0$.  It is assumed that the on-set of symptoms (coincident with infectiousness) is gradual after an incubation (exposed) phase. The values $\alpha_m$ and $\alpha_{sd}$  should be viewed as effective average values for a typical infected individual.
  \item Infectious individuals are removed due to hospitalization/isolation  at a rate $\mu_1(a)$ per day after disease age $a_{\mu1}$. This rate takes into account time lags in identifying presentation of symptoms.
  \item Infectious individuals are removed due to unreported mortality (independent of hospitalization)  at a rate 
$\mu_2(a)$ per day after disease age $a_{\mu2}$.  This rate allows for delays resultant from improperly handled corpses.
  \item Infectious individuals are (effectively) removed due to recovery (independent of hospitalization) at a rate $\mu_3(a)$ per day after disease age $a_{\mu3}$.  The values of $a_{\mu3}$ and $a_{\mu2}$ are significantly greater than the value of $a_{\mu1}$.
  \item Infectious individuals are removed  as a result of increased and earlier hospitalization at a rate $\mu_4(a,t)$ per day dependent on disease age $a$ and time $t \geq t_1$ during the course of the epidemic. The time $t_1$ is chosen based on case data indicating a lessening of the cumulative number of cases. We allow for a succession of times $t_1 <  t_2 < \dots$, when additional increased and earlier hospitalization occurs during the course of the epidemic.
\end{enumerate}

The formulas for $\mu_1(a)$, $\mu_2(a)$, $\mu_3(a)$, $\mu_4(a,t)$ are
\begin{equation}
\mu_1(a) = 0 \text{ if } a \leq a_{\mu1}, \, \mu_1(a) = \mu_{10} \text{ if } a > a_{\mu1}
\label{Eq2.1}
\end{equation}
\begin{equation}
\mu_2(a) = 0 \text{ if } a \leq a_{\mu2}, \, \mu_2(a) = \mu_{20} \text{ if } a > a_{\mu2}
\label{Eq2.2}
\end{equation}
\begin{equation}
\mu_3(a) = 0 \text{ if } a \leq a_{\mu3}, \, \mu_3(a) = \mu_{30} \text{ if } a > a_{\mu3}
\label{Eq2.3}
\end{equation}
\begin{equation}
\mu_4(a,t) = 0 \text{ if } t <t_1 \text{ or } a \leq a_{\mu4},  \, \mu_4(a,t) = \mu_{40} \text{ if }t  \geq t_1 \text{ and } a > a_{\mu4}.
\label{Eq2.4}
\end{equation}

The probabilities of removal by disease age $a$ for $\mu_j, \,  j=1,2,3,4$,  are
$$p \mu_j(a) = \int_0^a \mu_j(b) \exp\bigg(-\int_0^b(\mu_1(z)+\mu_2(z)+\mu_3(z)) dz \bigg) db, \,  j=1,2,3,$$ 
$$p \mu_4(a,t) = \int_0^a \mu_4(b,t) \exp\bigg(-\int_0^b(\mu_1(z)+\mu_2(z)+\mu_3(z)+\mu_4(z,t)) dz \bigg) db, \, t > t_1.$$ 

\section{The Equations of the Model}

The equations of the model are based on the disease age density $i(a,t)$, where $a$ is disease age and $t$ is time. Infected individuals have disease age $0$ at time of acquisition. The number of exposed (incubating) infected individuals at time $t$ is $E(t) = \int_0^{a_i} i(a,t) da$, where $a_i$ is the approximate disease age at which the infectious phase begins. The number of infectious individuals at time $t$ is $I(t) = \int_{a_i}^{a_r} i(a,t) da$, where $a_r$ is the approximate disease age at which infectiousness ends. The number of susceptible individuals at time $t$ is $S(t)$.
The model equations  are

\begin{equation}
\label{E1}
\frac{d}{dt} S(t)  = 
\ - \left( \int_{0}^{\infty} 
\alpha(a) i(a,t) da \right) S(t), t \geq 0
\end{equation}
\begin{equation}
\label{E2}
\frac{\partial}{\partial t} i(a,t) + \frac{\partial}{\partial a} 
i(a,t)  = -( \mu_1(a) + \mu_2(a) + \mu_3(a)+\mu_4(a,t))  i(a,t), t \geq 0, \, a \geq 0
\end{equation}
\begin{equation}
\label{E3}
i(0,t)  =  
\ \left(  \int_{0}^{\infty} 
\alpha (a) i(a,t) da \right) S(t), t \geq 0
\end{equation}
\begin{equation}
\label{E4}
\frac{d}{dt} R(t) =  \int_{0}^{\infty} 
( \mu_1(a) + \mu_2(a)+ \mu_3(a) + \mu_4(a,t))  i(a,t) da, t \geq 0
\end{equation}
\begin{equation}
\label{E5}
S(0)=S_0, \, i(a,0)=i_0 (a), a \geq 0, \, R(0)=0
\end{equation}

An analysis of the equations (\ref{E1})-(\ref{E5}) and the formulas below are given in \cite{Webb1}. The ultimate course of the epidemic satisfies $\lim_{t \rightarrow \infty} I(t) = 0$ and $\lim_{t \rightarrow \infty} S(t) > 0$. The reproduction number $R_0$ of the model before time $t_1$ is 
\begin{equation}
\label{R0noCT}
R_0 = S_0\, \int_0^{\infty}\alpha(a) \exp\bigg(-\int_0^a (\mu_1(b) + \mu_2(b) + \mu_3(b))db \bigg) da.
\end{equation}
and after time $t_1$  is
\begin{equation}
\label{R0CT}
R_0 = S(t_1)\, \int_0^{\infty}\alpha(a) \exp \bigg(-\int_0^a (\mu_1(b) + \mu_2(b)+ \mu_3(b)+\mu_4(b,t_1))db\bigg) da.
\end{equation}

The total cumulative number of cases at time $t$, both hospitalized (reported)  and non-hospitalized (unreported) is 
$$cumT(t) = S_0 - S(t) = \int_0^t i(0,s) ds.$$ 
The epidemic attack ratio at time $t, \, 0 \leq t \leq t_1,$ is 
$$AR(t) = \int_0^{t}i(0,t) dt/S_0.$$
The cumulative number of hospitalized cases at time $t, \, 0 \leq t \leq t_1,$  is 
$$cumH(t) = \int_0^t \int_0^{\infty} \mu_1(a) i(a,s) da \,  ds.$$ The cumulative number of unreported cases at time $t$ is $cumT(t) - cumH(t)$. The cumulative number of mortality cases (independent of hospitalization) at time $t, \, 0 \leq t \leq t_1,$  is 
$$cumM(t) =  \int_0^t \int_0^{\infty} \mu_2(a) i(a,s) da \, ds.$$
The cumulative number of recovered cases (independent of hospitalization) at time $t, \, 0 \leq t \leq t_1,$  is 
$$cumR(t) = \int_0^t \int_0^{\infty}\mu_3(a) i(a,s) da ds.$$ Similar formulas can be obtained for $AR(t)$, $cumH(t)$, $cumM(t)$, and $cumR(t)$ for $t \geq t_1$.

\section{Simulations of the Model for the Ebola Epidemic in Sierra Leone}

The model (\ref{E1})-(\ref{E5}) is simulated for the 2014-2015 Ebola epidemic in Sierra Leone. For this simulation the epidemic is divided into 3 phases: May 27, 2014 to September 1, 2014, September 1 to December 14, 2014, and December 14, 2014 forward in time. The simulation phase time values are $t_1 = $ September 1, 2014 and $t_2 =$ December 14, 2014, where each phase represents an increased removal of cases due to a greater number of cases hospitalized, with earlier hospitalization. The removal rates $\mu_1(a)$, $\mu_2(a)$, $\mu_3(a)$ hold for the first, second, and third phases ($t \geq 0$). The removal rate  $\mu_4(a,t)$ holds for the second and third phases ($t \geq t_1$). Additionally, there is the removal rate $\mu_5(a,t)$ holding for the third phase $t \geq t_2$ having the form
\begin{equation}
\mu_5(a,t) = 0 \text{ if } t <t_2 \text{ or } a \leq a_{\mu5},  \, \mu_5(a,t) = \mu_{50} \text{ if }t  \geq t_2 \text{ and } a > a_{\mu5}.
\label{Eq4.1}
\end{equation}

\begin{figure}
\begin{center}
\includegraphics[width=13.0cm,height=7.0cm]{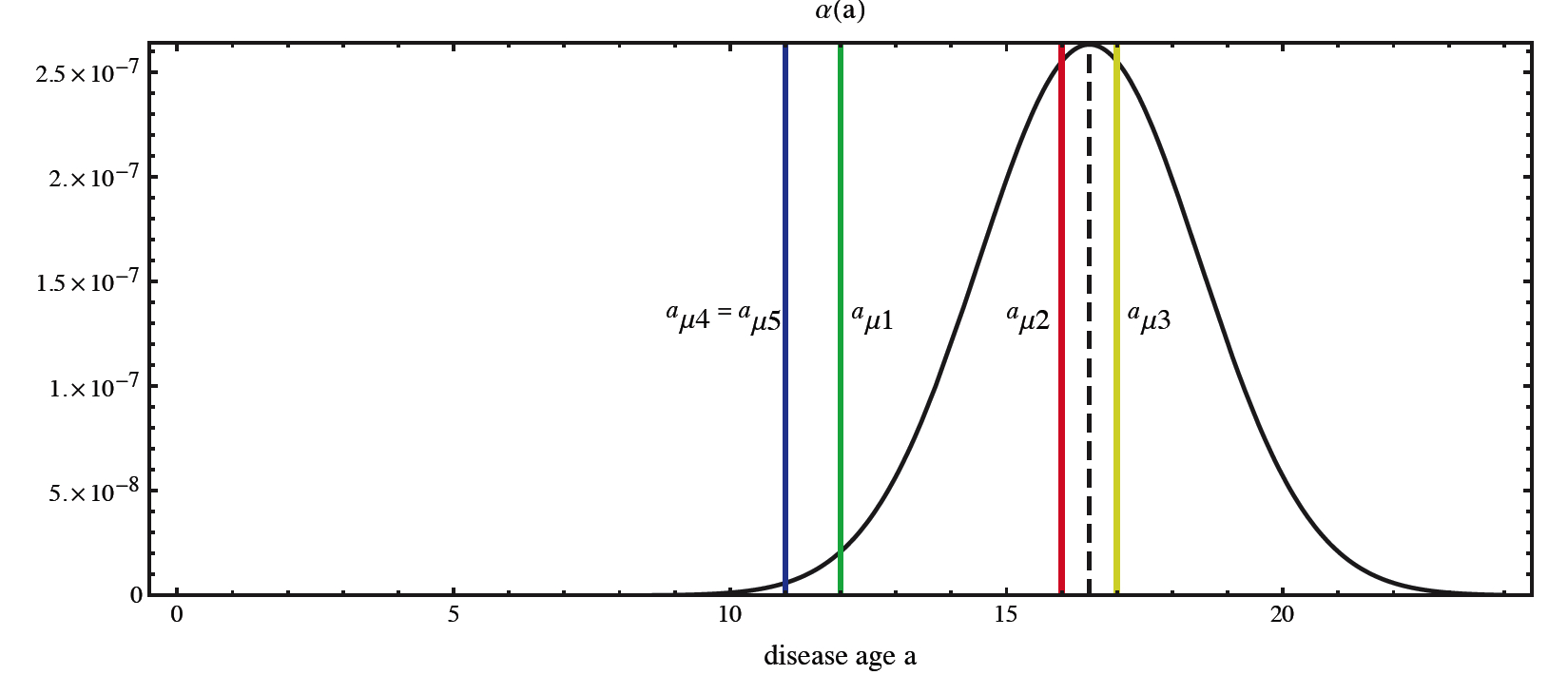}
\caption{\label{fig1}The disease age infection transmission function for Sierra Leone $\alpha(a)$ with $\alpha_0 =  1.27 \times 10^{-6}$, $\alpha_m = 16.5$, $\alpha_{sd} = 2.0$, and the removal rate start ages $a_{\mu1} = 12$, $a_{\mu2} = 16, \, a_{\mu3} =17$, \, $a_{\mu4} = 11$, $a_{\mu5} = 11$. The infectious phase begins at $\approx 10$ days and ends at 
$\approx 23$ days.}
\end{center}
\end{figure}

\begin{figure}
\begin{center}
\includegraphics[width=14.0cm,height=8.0cm]{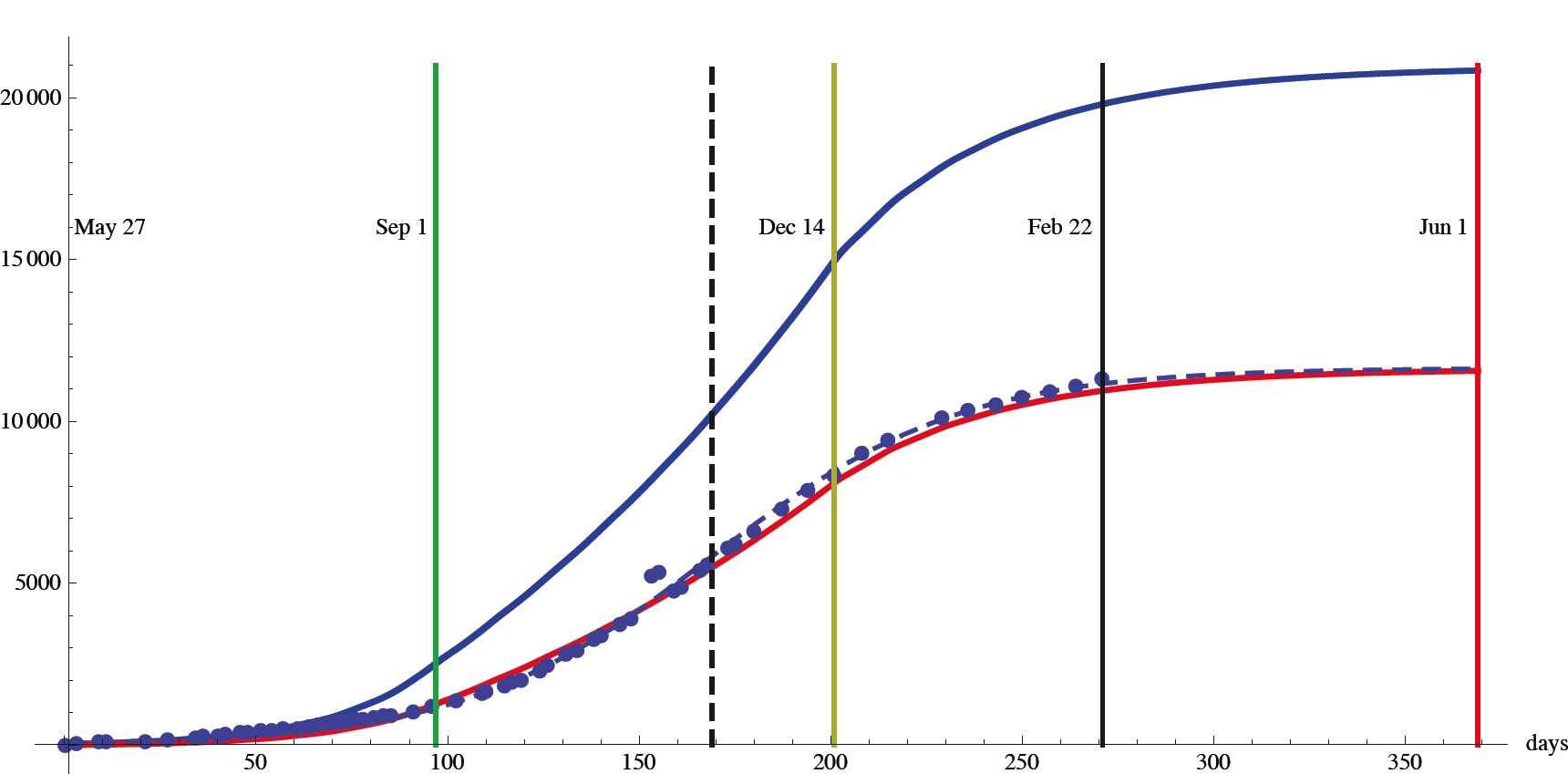}
\caption{\label{fig2} Simulation of the Ebola epidemic in Sierra Leone with the parameters in Table 1. The dots are  reported cumulative case data \cite{WHO1}. The red graph is reported cumulative cases from the simulation. The blue graph is the total cumulative cases from the simulation (both reported and unreported). $R_0 = 1.77$ between May 27, 2014 and September 1, 2014.  $R_0 = 1.10$ between September 1 and December 14, 2014.  $R_0 = 0.72$ between December 14, 2014 and forward in time.}
\end{center}
\end{figure}

\begin{figure}
\begin{center}
\includegraphics[width=13.0cm,height=9.0cm]{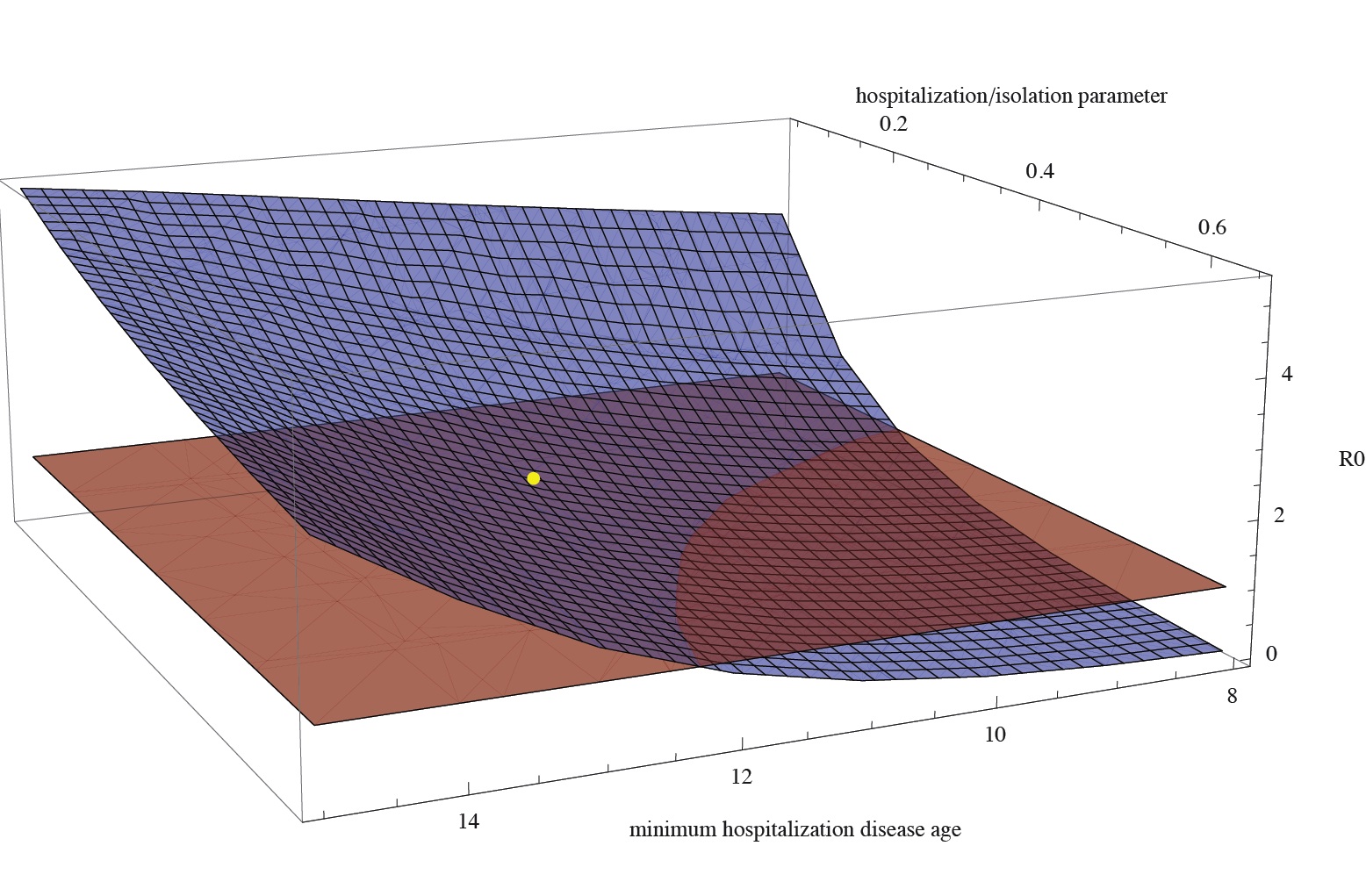}
\caption{\label{fig3} $R_0$ for Sierra Leone during the first phase between May 27, 2014 and September 1, 2014 as a function of $a_{\mu1}$ (the earliest disease age for hospitalization) and $\mu_{10}$ (the hospitalization parameter) with all other parameters as in Table 1.  The blue surface is $R_0 = R_0(a_{\mu1},\mu_{10})$ and the red plane is $R_0 \equiv 1.0$. The yellow dot corresponds to the Table 1 values $a_{\mu1} = 12$ days, $\mu_{10} = 0.35$ per day, and $R_0 = 1.77$.}
\end{center}
\end{figure}

The parameters for the simulation are given in Table 1. The model parameters are chosen by fitting the model solution to cumulative reported case data from WHO situation reports \cite{WHO1}. The graph of the disease age dependent transmission function $\alpha(a)$ is given in Figure \ref{fig1}. The simulation of the model is illustrated in Figure \ref{fig2}. The phase times $t_1$ and $t_2$ are chosen based on fitting this data to a solution of the logistic equation. The concavity change in the data corresponds to the transition over time of $R_0$ from $ > 1$ to $ < 1$, which corresponds in turn to the concavity change in the fitted logistic equation solution. The logistic equation and its solution are
\begin{equation}
P^{\prime}(t) = r P(t) ( 1 - P(t) / C), \, \, \, \, P(t) = \frac{e^{r t} C P(0)}{C-P(0) + e^{r t} P(0)}
\label{EqLOG}
\end{equation}
The dashed graph in Figure \ref{fig2} is the solution of (\ref{EqLOG}) with $r = 0.031$, $C = 11,650$, and $P(0) = 63$. The dashed vertical line in Figure \ref{fig2} is the concavity change in $P(t)$ at  $\bar{t} \approx 169$, $P(\bar{t}) = C/2$. In Figure \ref{fig2} the simulation is projected forward in time to June 1, 2015. The dependence of $R_0$ during the first phase between May 27, 2014 and September 1, 2014 as a function of $a_{\mu1}$ (the earliest disease age for hospitalization in the first phase) and $\mu_{10}$ (the hospitalization rate after disease  age $a_{\mu1}$ in the first phase) is illustrated in Figure \ref{fig3} (all other parameters are as in Table 1). Figure \ref{fig3} demonstrates the importance of smaller values of $a_{\mu1}$ and larger values of $\mu_{10}$ for mitigation of the epidemic.  The simulation for Sierra Leone is summarized in Table 2.

\section{Simulations of the Model for the Ebola Epidemic in Guinea}

\begin{figure}
\begin{center}
\includegraphics[width=13.0cm,height=8.0cm]{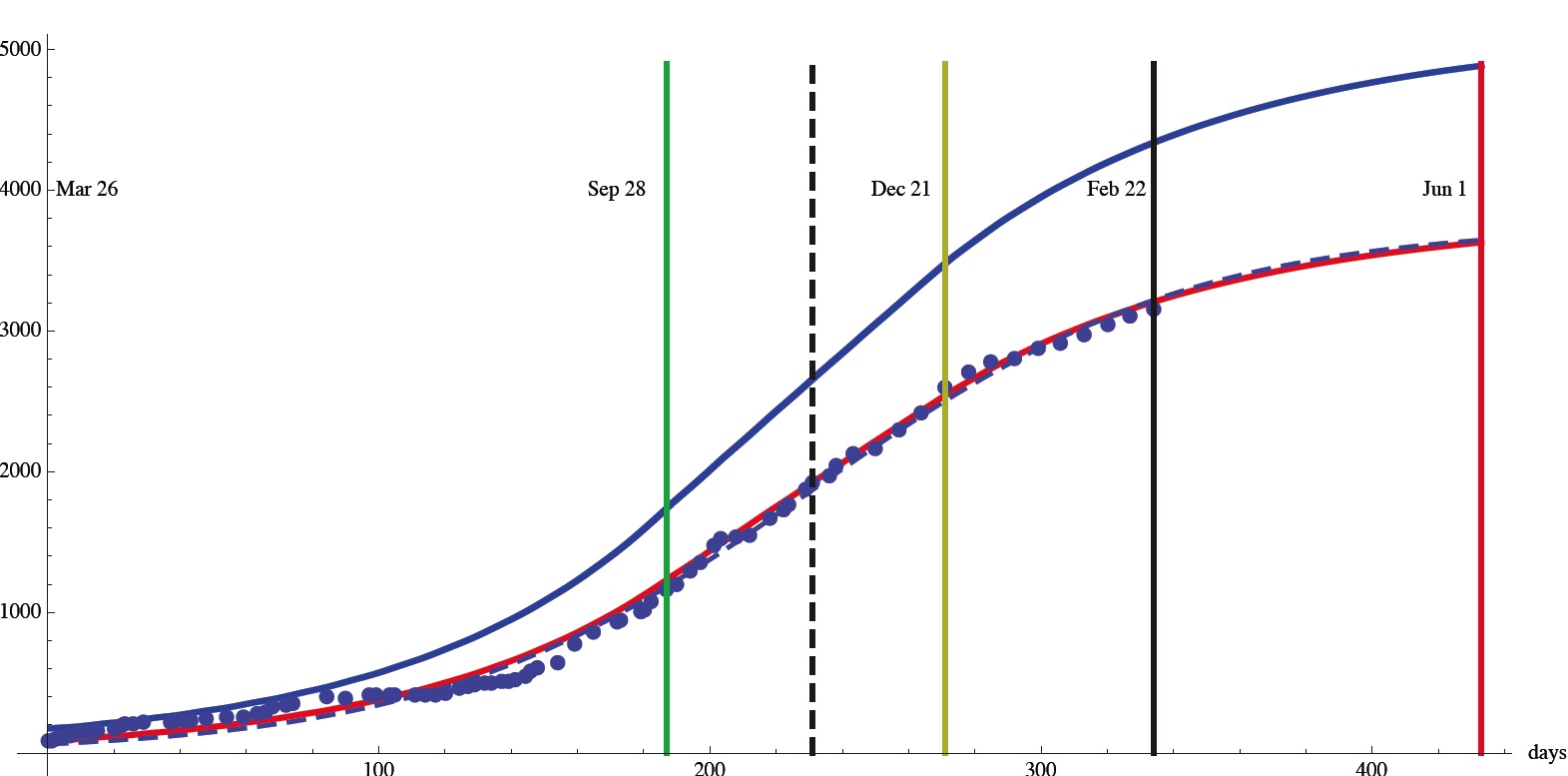}
\caption{\label{fig4} Simulation of the Ebola epidemic in Guinea. The dots are  reported cumulative case data \cite{WHO1}. The red graph is reported cumulative cases from the simulation. The blue graph is the total cumulative cases from the simulation (both reported and unreported). $R_0 = 1.22$ between March 26, 2014 and September 28, 2014.  $R_0 = 1.00$ between September 28 and December 21, 2014.  $R_0 = 0.85$ between December 21, 2014 and forward in time.}
\end{center}
\end{figure}

\begin{figure}
\begin{center}
\includegraphics[width=13.0cm,height=9.0cm]{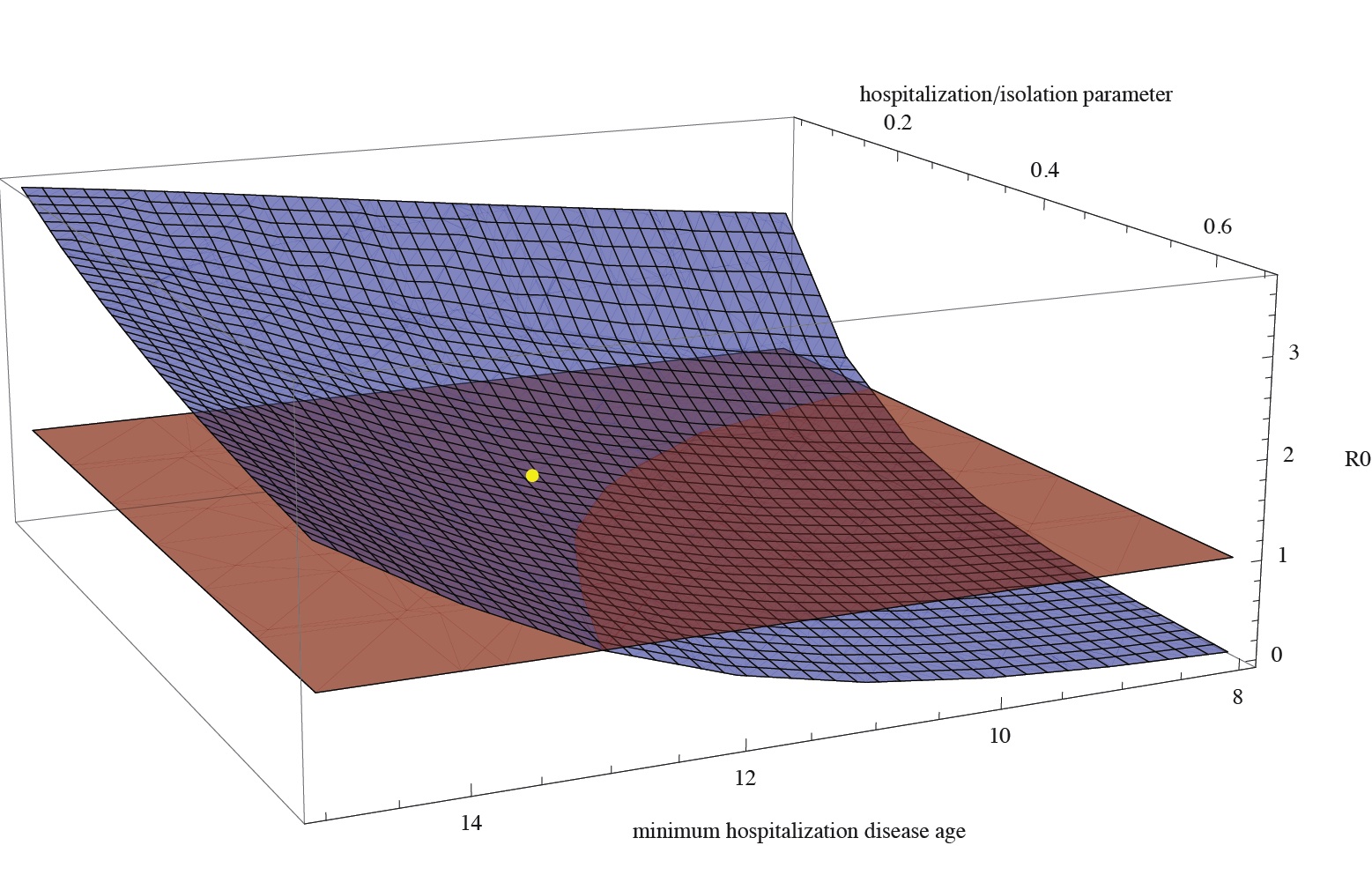}
\caption{\label{fig5}$R_0$ for Guinea during the first phase between March 26, 2014 and September 28, 2014 as a function of $a_{\mu1}$ (the earliest disease age for hospitalization) and $\mu_{10}$ (the hospitalization parameter) with all other parameters as in Figure \ref{fig4}.  The blue surface is $R_0 = R_0(a_{\mu1},\mu_{10})$ and the red plane is $R_0 \equiv 1.0$. The yellow dot corresponds to the values $a_{\mu1} = 12$ days, $\mu_{10} = 0.35$, and $R_0 = 1.22$.}
\end{center}
\end{figure}

The model (\ref{E1})-(\ref{E5}) is simulated for the 2014-2015 Ebola epidemic in Guinea. The parameters for the model simulation are chosen as for Sierra Leone, based on WHO cumulative reported case data \cite{WHO1}. For this simulation the epidemic is divided into 3 phases: March 26, 2014 to September 28, 2014, September 28, 2014 to December 21, 2014, and December 21, 2014 forward in time. The simulation phase time values are $t_1 = $ September 28, 2014 and $t_2 =$ December 21, 2014. The parameterization for the simulation is the following: $S_0 = 10.6 \times 10^6$; $i_0(a)$ is the same as for Sierra Leone (Table 1); the disease age dependent transmission rate $\alpha(a)$ is the same as for Sierra Leone (Table 1) except $\alpha_0 = 0.49 \times 10^{-6}$; $a_{\mu1}$, $a_{\mu2}$, $a_{\mu3}$, $a_{\mu4}$, $a_{\mu5}$ are the same as for Sierra Leone (Table 1); the removal rate parameters are 
$\mu_1(a) = 0.35$, $\mu_2(a) = 0.22$, $\mu_3(a) = 0.30$, $\mu_4(a,t) =  .053$, $\mu_5(a,t) = .045$.
The logistic equation parameters in (\ref{EqLOG}) are $r = 0.175$, $C = 3,752$, and $P(0) = 65$, with concavity change at $\bar{t } = 231$, $P(\bar{t}) = C/2$.
The simulation of the model is illustrated in Figure \ref{fig4}.  In Figure \ref{fig4} the simulation is projected forward in time to June 1, 2015. The dependence of $R_0$ during the first phase between Mar 26, 2014 and September 28, 2014 as function of $a_{\mu1}$ (the earliest disease age for hospitalization in the first phase) and $\mu_{10}$ (the parameter for hospitalizationafter age disease  $a_{\mu1}$ in the first phase) is illustrated in Figure \ref{fig5}. The simulation for Guinea is summarized in Table 2.

\section{Discussion}

We have presented a model of the 2014-2015 Ebola epidemics in Sierra Leone and Guinea structured by  the disease age of infected individuals. The incorporation of disease age allows refinement of the disease transmission rate based on level of infectiousness.  For Ebola the level of infectiousness is correlated to the level of symptoms, which begin gradually over a period of several days.  The incorporation of disease age also allows refinement of removal rates due to hospitalization, mortality, and recovery correlated to disease age. Consequently, the significant advantage in hospitalization/isolation of infected individuals as soon as possible after presentation of symptoms is emphasized.

Our model simulations are fitted to WHO cumulative reported case data \cite{WHO1} through February 22, 2015 and projected forward to June 1, 2015. The data indicate that the Ebola epidemics have subsided and will further subside to effective elimination in 2015 for these two countries. The reasons for the extinction of these Ebola epidemics are complex, but increased identification and isolation of infectious cases played a major role. Our model quantifies the significance of this increased identification/isolation. One way to quantify an epidemic is the
epidemic net reproduction number $R_0$, which is dynamic and evolves with changes in public health interventions and societal behavior. In our simulations we modeled this  $R_0$ dynamic with two changes in hospitalization/isolation rates in two successive time periods. Our model simulations indicate that this removal of infectious individuals is key to elimination of the epidemic. This removal is quantified in our model in two ways: (1) with respect to the  rate of infectious individuals hospitalized per day, and (2) with respect to an earlier disease age of the infectious individuals hospitalized. Both of these considerations can be influenced by public health policies and public awareness.

\newpage

\newpage

\newpage

\newpage

\begin{figure}[ht]
\begin{center}
\includegraphics[width=15.0cm,height=15.0cm]{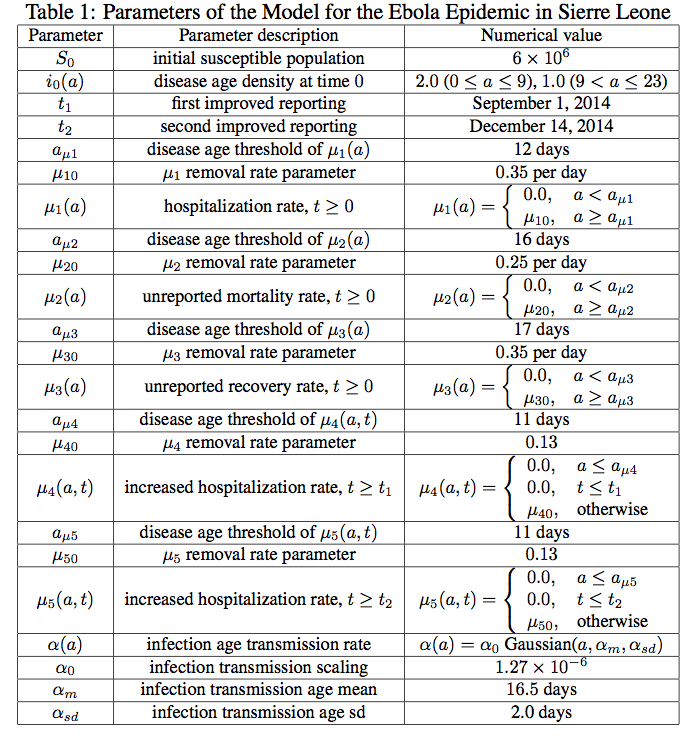}
\end{center}
\end{figure}

\newpage

\begin{figure}[ht]
\begin{center}
\includegraphics[width=13.0cm,height=9.0cm]{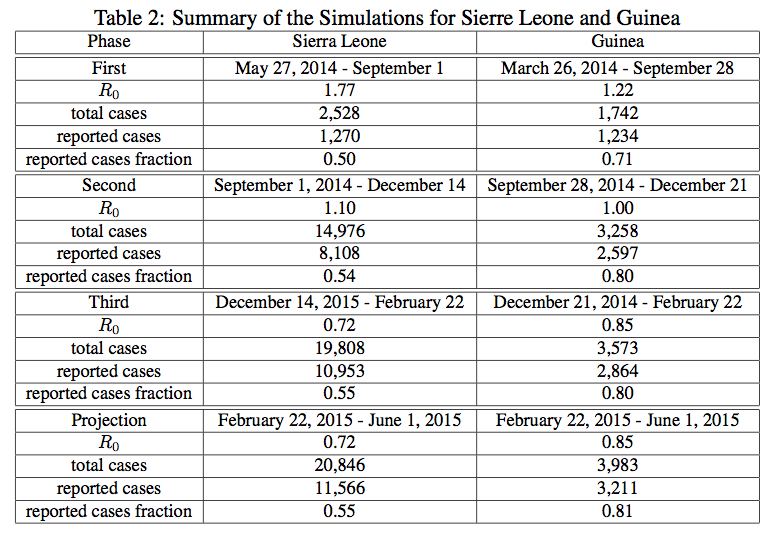}
\end{center}
\end{figure}

\end{document}